\documentclass[mathleft
]{an}
\usepackage{graphicx}
\usepackage{times}
\def\vsin{\hbox{$v \sin i$}}

\def\em{\it}

\def\msun{\hbox{$M_\odot$}}

\begin{document}

\Pagespan{789}{}
\Yearpublication{2006}%
\Yearsubmission{2005}%
\Month{11}%
\Volume{999}%
\Issue{88}%

\title{Direct observation of magnetic cycles in Sun-like stars}

\author{A. Morgenthaler\inst{1,2}\fnmsep\thanks{Corresponding author:
  \email{amorgent@ast.obs-mip.fr}\newline}
\and  P. Petit\inst{1,2}
\and  J. Morin\inst{3}
\and  M. Auri\`ere\inst{1,2}
\and  B. Dintrans\inst{1,2}
\and  R. Konstantinova-Antova\inst{4}
\and  S. Marsden\inst{5}
}
\titlerunning{Direct observation of magnetic cycles in Sun-like stars}
\authorrunning{A. Morgenthaler et al.}
\institute{
Universit\'e de Toulouse, UPS-OMP, Institut de Recherche en Astrophysique et Plan\'etologie, Toulouse, France
\and 
CNRS, Institut de Recherche en Astrophysique et Plan\'etologie, 14 Avenue Edouard Belin, F-31400, Toulouse, France
\and
Dublin Institute for Advanced Studies, School of Cosmic Physics, 31 Fitzwilliam Place, Dublin 2, Ireland
\and
Institute of Astronomy, Bulgarian Academy of Sciences, 72 Tsarigradsko shose, 1784 Sofia, Bulgaria\\
\and
Centre for Astronomy, School of Engineering and Physical Sciences, James Cook University, Townsville 4811, Australia
}

\received{}
\accepted{}
\publonline{}


\abstract{A sample of 19 solar-type stars, probing masses between 0.6 and 1.4 solar mass and rotation periods between 3.4 and 43 days, was regularly observed using the NARVAL spectropolarimeter at Telescope Bernard Lyot (Pic du Midi, France) between 2007 and 2011. The Zeeman-Doppler Imaging technique is employed to reconstruct the large-scale photospheric magnetic field structure of the targets and investigate its long-term temporal evolution. We present here the first results of this project with the observation of short magnetic cycles in several stars, showing up a succession of polarity reversals over the timespan of our monitoring. Preliminary trends suggest that short cycles are more frequent for stellar periods below a dozen days and for stellar masses above about one solar mass. The cycles lengths unveiled by the direct tracking of polarity switches are significantly shorter than those derived from previous studies based on chromospheric activity monitoring, suggesting the coexistence of several magnetic timescales in a same star.}

\maketitle

\section{Introduction}
Sun-like stars are characterized by convective envelopes in which large-scale plasma flows (related, in particular, to radial and latitudinal differential rotation and to the Coriolis force) are able to trigger a global dynamo (Parker 1955). This continuous generation of a large-scale field is related to surface variability affecting a wide range of temporal and spatial scales, including quasi-periodic polarity reversals associated to magnetic cycles. Recent numerical models, in particular global MHD simulations, are able to mimic some characteristics of this cyclic behaviour for Sun-like stars (Ghizaru et al. 2010, Brown et al. 2011). 

This magnetic activity is observable through many observational proxies, from the photosphere (e.g. broad-band visible photometry) to the corona (X-ray and radio emissions). The long-term monitoring of chromospheric activity of tens of solar-type stars, carried out at Mount Wilson since 1965 (Wilson 1978) has allowed for the detection of periodic variations in a number of objects (Baliunas et al. 1995). Cyclic patterns were observed with a variety of cycle lengths, as well as seemingly more erratic activity fluctuations in other objects (in particular young stars) or no detectable activity at all in some others. More recently, asteroseismology was demonstrated its ability to investigate magnetic variability through variations of the p-mode amplitudes and frequencies (Garcia et al. 2010) with the detection of variations of these quantities over a few tens of days, presumably linked to magnetic activity, for the F5V star HD 49933.

In addition to this wealth of indirect indicators of stellar activity, spectropolarimetry now enables us to perform direct measurements of surface magnetic fields and follow the long-term temporal evolution of large-scale magnetic geometries. So far, it allowed the observation in Sun-like stars of one global polarity switch (Petit et al. 2009) and of a full magnetic cycle (Fares et al. 2009).

Our aim is to study the long-term variations of the magnetic field properties of a sample of solar-type stars, using both direct and indirect measurements. Our observed sample includes 19 FGK-type stars on the main sequence, monitored since 2007. We probe here stellar masses between 0.6 and 1.4\msun, and rotation periods between 3.4 and 43 days.

After a brief description of the instrumental setup, data reduction and multi-line extraction of Zeeman signatures, we explain the reconstruction technique of the large-scale topology of the stars, and the computing of a chromospheric activity indicator. We then highlight three representative examples of different types of variability observed in our sample. We finally discuss the results derived from our measurements.

\section{Instrumental setup, data reduction, and extraction of Zeeman signatures}

We use data from the NARVAL spectropolarimeter (Auri\`ere 2003), installed at Telescope Bernard Lyot\footnote{The Bernard Lyot Telescope is operated by the Institut National des Sciences de l'Univers of the Centre National de la Recherche Scientifique of France.} (Pic du Midi, France). The instrumental setup is strictly identical to the one described by Petit et al. (2008). The spectrograph unit of NARVAL benefits from a spectral resolution of 65,000 and covers the whole wavelength domain from near-ultraviolet (370 nm) to near-infrared (1,000 nm). Thanks to the polarimetric module, NARVAL can provide intensity, circularly or linearly polarized spectra. In the present study, we restrict the measurements to Stokes I and V. 

\begin{figure}
\centering
\includegraphics[scale=0.5]{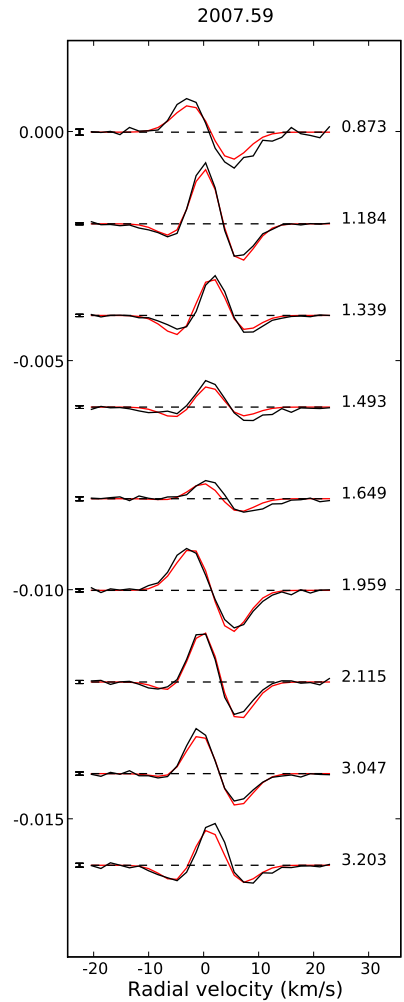}
\caption{Normalized Stokes V profiles of $\xi$ Boo A for the summer of 2007, after correction of the mean radial velocity of the star. Black line represent the data and red lines correspond to synthetic profiles of our magnetic model. Successive profiles are shifted vertically for display clarity. Rotational phases of observations are indicated in the right part of the plot and errors bars are illustrated on the left of each profile.}
\label{fig:stokes}
\end{figure}

The circularly polarized spectra allow the detection of large-scale photospheric magnetic fields, thanks to the Zeeman effect. However, when observing cool dwarfs, the sig\-nal-to-noise ratio of circularly polarized spectra pro\-du\-ced by NARVAL is not sufficiently high to reach the detection threshold of typical Zeeman signatures (which amplitude does not exceed $10^{-4}I_c$ for low-activity stars, where $I_c$ is the continuum intensity). To solve this problem, we calculate from the reduced spectrum a single, cross-cor\-re\-lated photospheric line profile using the Least-Squares-De\-con\-vo\-lution (LSD) multi-line tech\-ni\-que (detailed by Donati et al. 1997 and Kochukhov et al. 2010). Thanks to the large number of available photospheric lines in cool stars (several thousands in the spectral domain of NARVAL), the noise level is reduced by a factor of about 30 with respect to the initial spectrum. As an illustration, Fig. \ref{fig:stokes} shows the resulting LSD signatures for successive observations of the Sun-like star $\xi$ Boo A.

\section{Magnetic mapping and chromospheric emission}

The Stokes I and Stokes V LSD profiles allow the derivation of various quantities to study the temporal variations of the magnetic field properties. Here we focus on the reconstruction of the surface distribution of the magnetic vector and on the computation of a chromospheric activity index.

\subsection{Magnetic maps}

To reconstruct the surface magnetic geometry of the stars, we use Zeeman-Doppler Imaging (ZDI). This tomographic inversion technique is based on the modelling of the rotational modulation of the circularly polarized signal (Semel 1989). The time series of polarized signatures are iteratively compared to artificial profiles corresponding to a synthetic magnetic geometry, until a good fit is obtained between the model and the observations (Donati \& Brown 1997, Donati et al. 2006). Thus, ZDI enables to recover, to some extent, the location of magnetic regions, as well as the strength and orientation of the magnetic vector in magnetic spots. The application of this technique to cool stars with low \vsin\ and moderate to low magnetic activity is described by Petit et al. (2008). In this case, ZDI is only sensitive to low-order field components, contrary to the chromospheric flux which includes also the contribution of smaller scale magnetic elements.

The resulting maps for the three stars presented here are illustrated in Fig. \ref{fig:hd78366}, \ref{fig:hd190771} and \ref{fig:ksiboo}. 

\subsection{$N_{CaII H}$-index}

From the Stokes I profiles, we construct an index to quantify the chromospheric emission changes in the CaII H line. The complete pipeline of the computation is described in details in a forthcoming paper (Morgenthaler et al., submitted). We follow the methods of Duncan et al. (1991) and Wright et al. (2004), who define indexes based on Mount Wilson observations, and we calculate a $N_{CaIIH}$-index for our NARVAL observations. To help comparing our chromospheric estimate with older studies, we calibrated the index against the values derived at Mount Wilson. The $N_{CaIIH}$-indexes we obtained for HD 78366, HD 190771 and $\xi$ Boo A are detailed in Tab. \ref{tab:indice}.

\section{Results}

\begin{table*}
\centering
\caption[]{Fundamental parameters of HD 78366, HD 190771 and $\xi$ Boo A. $T_{eff}$, mass, radius and \vsin\ are taken from Valenti \& Fischer (2005), Fernandes et al. (1998), Gray (1984), Petit et al. (2005). The rotation periods and inclination angles are derived from ZDI and from Toner \& Gray (1988).}
\begin{tabular}{ccccccc}
\hline
Star & $T_{eff}$ & Mass         & Radius & $v$sin$i$ & $P^{eq}_{rot}$ & inclination \\
     & (K)      & ($M_{\odot}$) & ($R_{\odot}$) & (km/s)    & (d)           & (degrees) \\
\hline
HD 78366 & $6014 \pm 50$ & $1.34 \pm 0.13$ & $1.03 \pm 0.02$ & $3.9 \pm 0.5$ & $11.4 \pm 0.1$ & $60 \pm 15$ \\
HD 190771  & $5834 \pm 50$ & $0.96 \pm 0.13$ & $0.98 \pm 0.02$ & $4.3 \pm 0.5$ & $8.8 \pm 0.1$ & $50 \pm 10$ \\
$\xi$ Boo A & $5551 \pm 20$ & $0.86 \pm 0.07$ & $0.80 \pm 0.03$ & $3.0 \pm 0.4$ & $6.43$ & $28 \pm 5$ \\
\hline
\end{tabular}\\
\label{tab:param}
\end{table*}

Since the monitoring began a few years ago, long-term chan\-ges in the magnetic properties become observable in some of our targets. Both the magnetic quantities derived from ZDI and the chromospheric index exhibit temporal fluctuations over a wide range of timescales, due to rotational modulation and longer-term magnetic trends. Three representative examples of the different kinds of stellar variability we observed is described hereafter.

\begin{table}
\centering
\caption[]{Chromospheric activity indices for HD 78366, HD 190771 and $\xi$ Boo A for each corresponding set of observations.}
\begin{tabular}{ccc}
\hline
Star & Set of obs. & $N_{CaIIH}$ \\
\hline
\hline
HD 78366 & 2008.09 & $0.273 \pm 0.004$ \\
         & 2010.04 & $0.291 \pm 0.006$ \\
         & 2011.08 & $0.278 \pm 0.003$ \\
\hline
HD 190771 & 2007.59 & $0.335 \pm 0.006$ \\
          & 2008.67 & $0.338 \pm 0.011$ \\
          & 2009.47 & $0.337 \pm 0.007$ \\
          & 2010.50 & $0.345 \pm 0.006$ \\
\hline
$\xi$ Boo A & 2007.59 & $0.443 \pm 0.008$ \\
            & 2008.09 & $0.420 \pm 0.008$ \\
            & 2010.48 & $0.403 \pm 0.006$ \\
            & 2010.59 & $0.402 \pm 0.011$ \\
\hline
\end{tabular}\\
\label{tab:indice}
\end{table}

\subsection{Short magnetic cycle : HD 78366}

\begin{figure*}
\centering
\mbox{
\includegraphics[width=5.5cm]{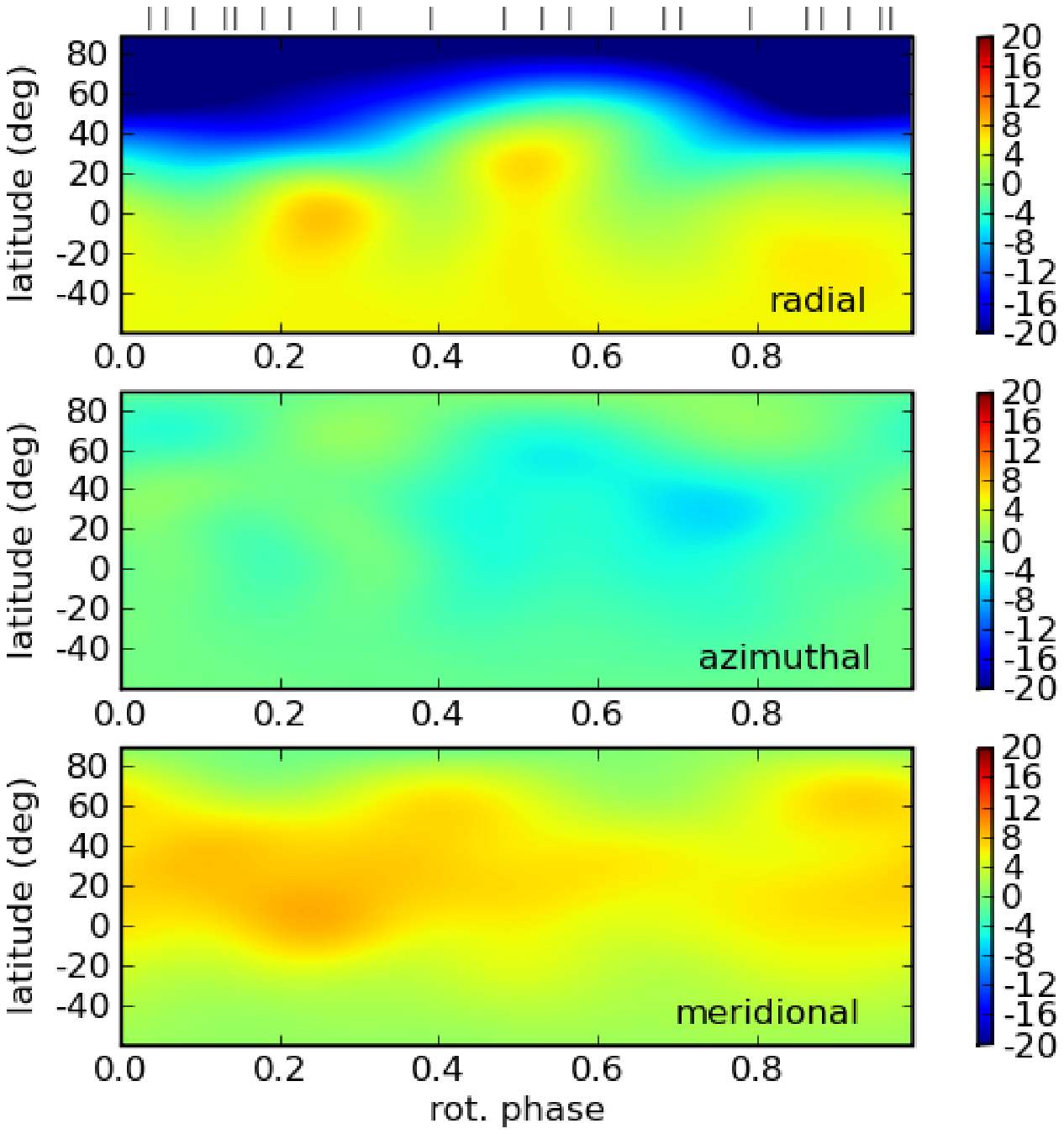}
\includegraphics[width=5.5cm]{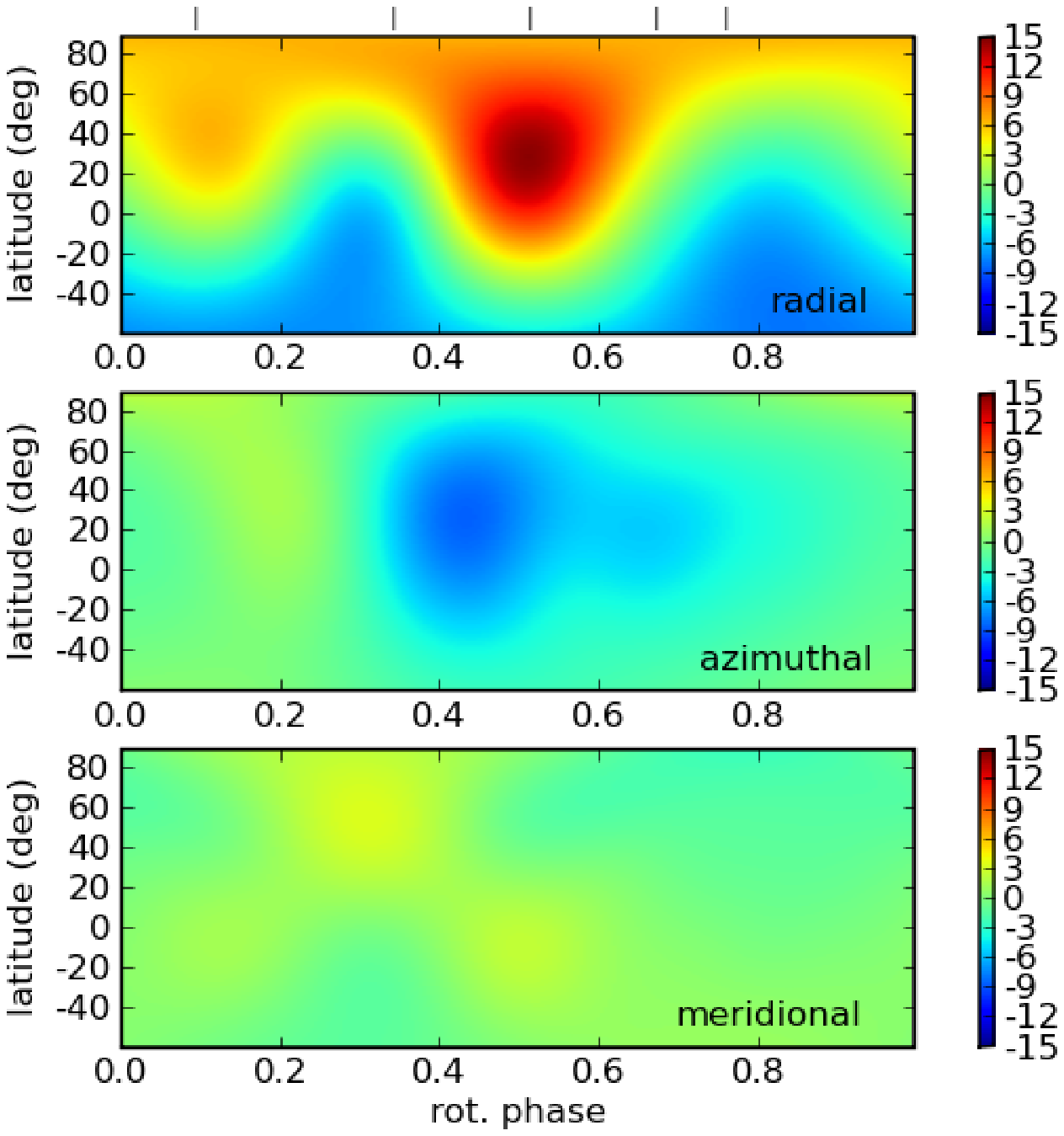}
\includegraphics[width=5.5cm]{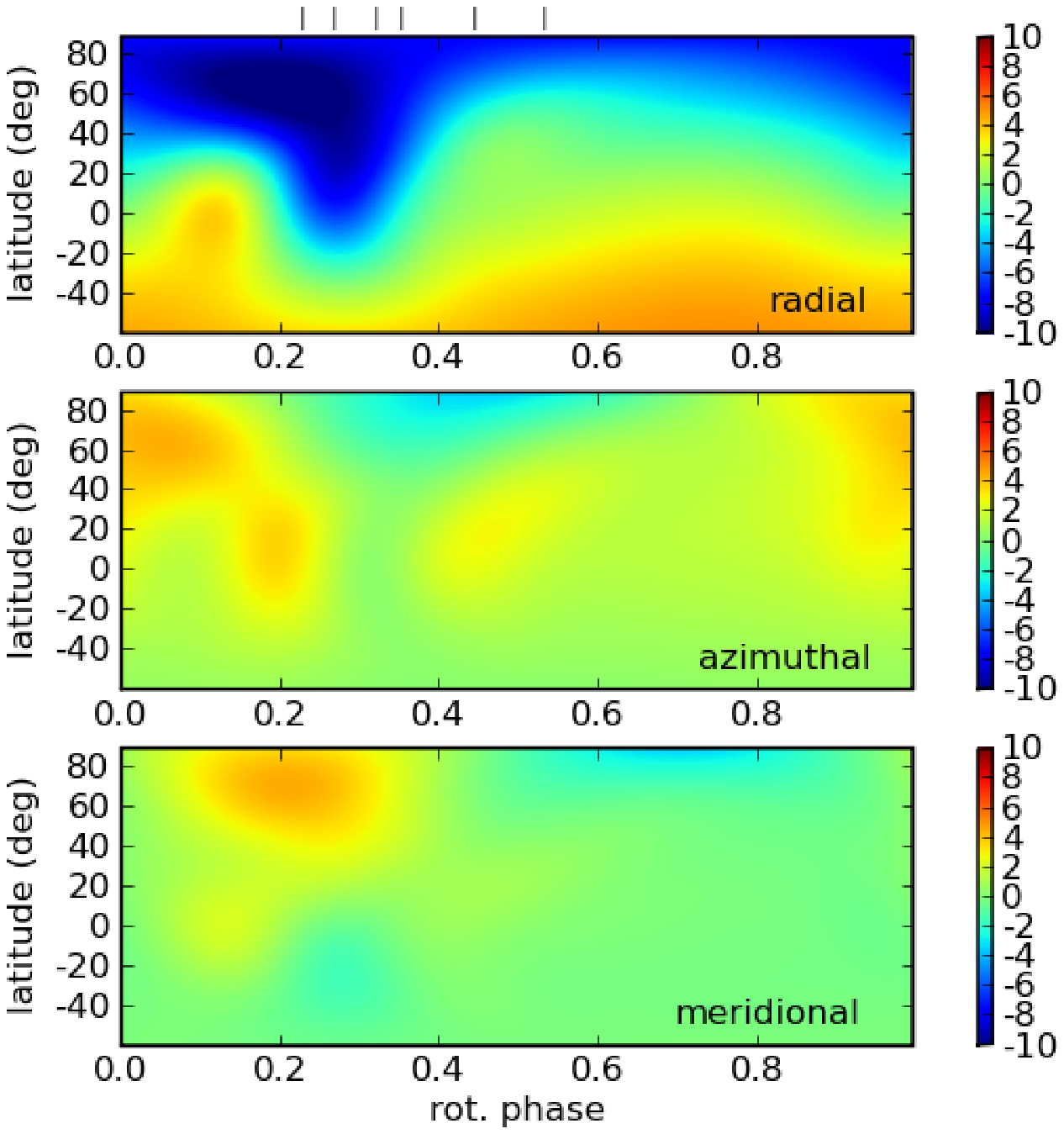}
}
\caption{Magnetic maps of HD 78366, derived from 2008.09, 2010.04 and 2011.08 observations (from left to right). For each data set, the 3 charts illustrate the field projection onto one axis of the spherical coordinate frame with, from top to bottom, the radial, azimuthal, and meridional field components. The magnetic field strength is expressed in Gauss.}
\label{fig:hd78366}
\end{figure*}

A simple type of variability is observed for HD 78366. This target is slightly more massive than the Sun and rotates about two times faster (Tab. \ref{tab:param}). The data sets of this object are collected over three distant seasons. The corresponding magnetic maps are shown in Fig. \ref{fig:hd78366}. We observe two polarity switches, especially visible in the polar area of the radial field component, which is of negative polarity in 2008.09, positive in 2010.04 (and associated at that time with a more complex magnetic field geometry), and negative a\-gain in 2011.08. After the two observed polarity reversals, the magnetic field retrieves its initial configuration. Assuming that the magnetic variability of this star is not much faster than the temporal sampling imposed by the right ascension of the star (visible only during winters), this first time-series suggests that HD 78366 may obey to a magnetic cycle of about three years. We note that the chromospheric activity indicator $N_{CaII H}$ (Tab. \ref{tab:indice}) seems to increase with the complexity of the large-scale magnetic field, with a clear maximum in 2010.04.

\subsection{Fast polarity reversals : HD 190771}

\begin{figure*}
\centering
\mbox{
\includegraphics[width=6cm]{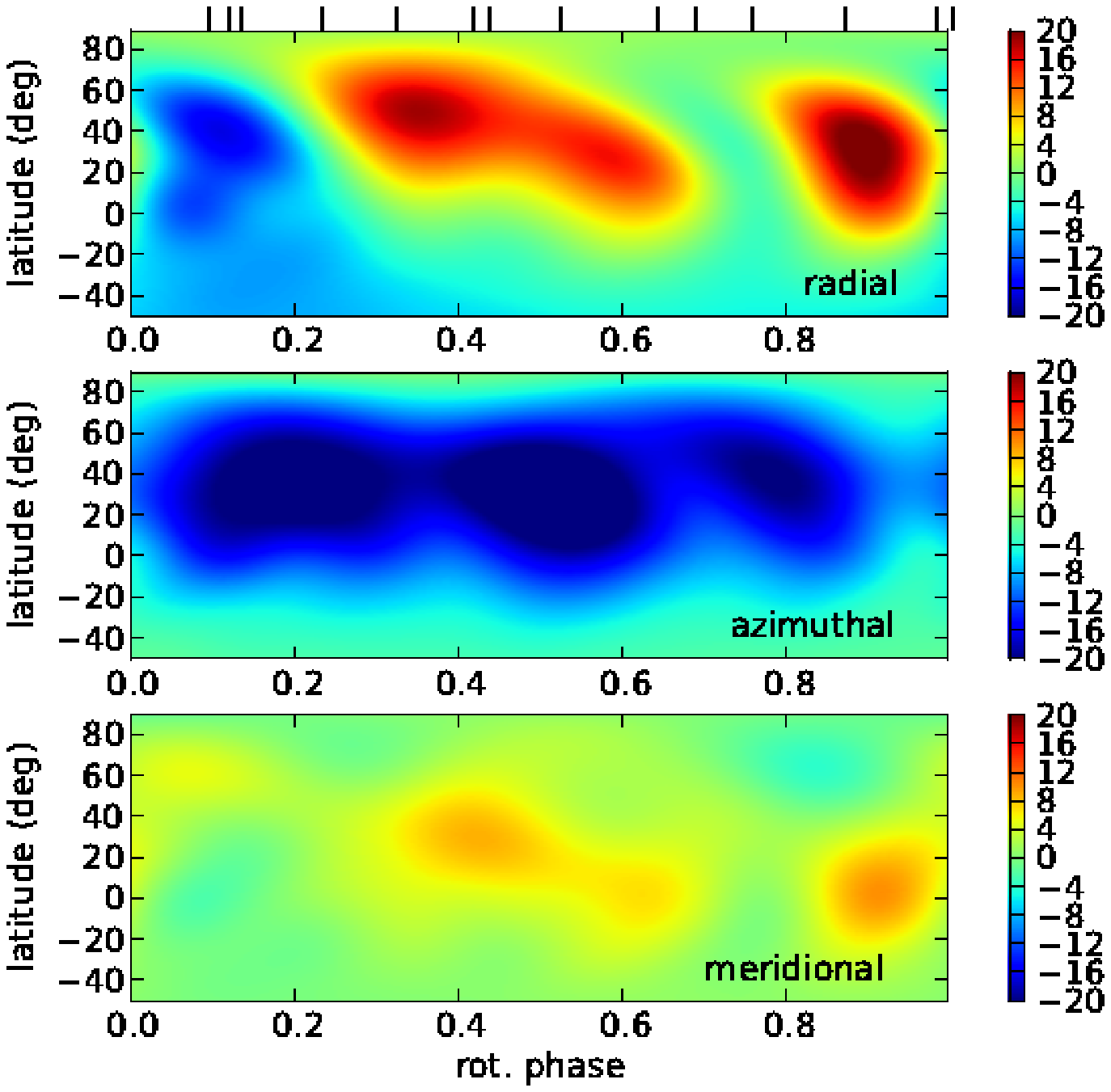}
\includegraphics[width=6cm]{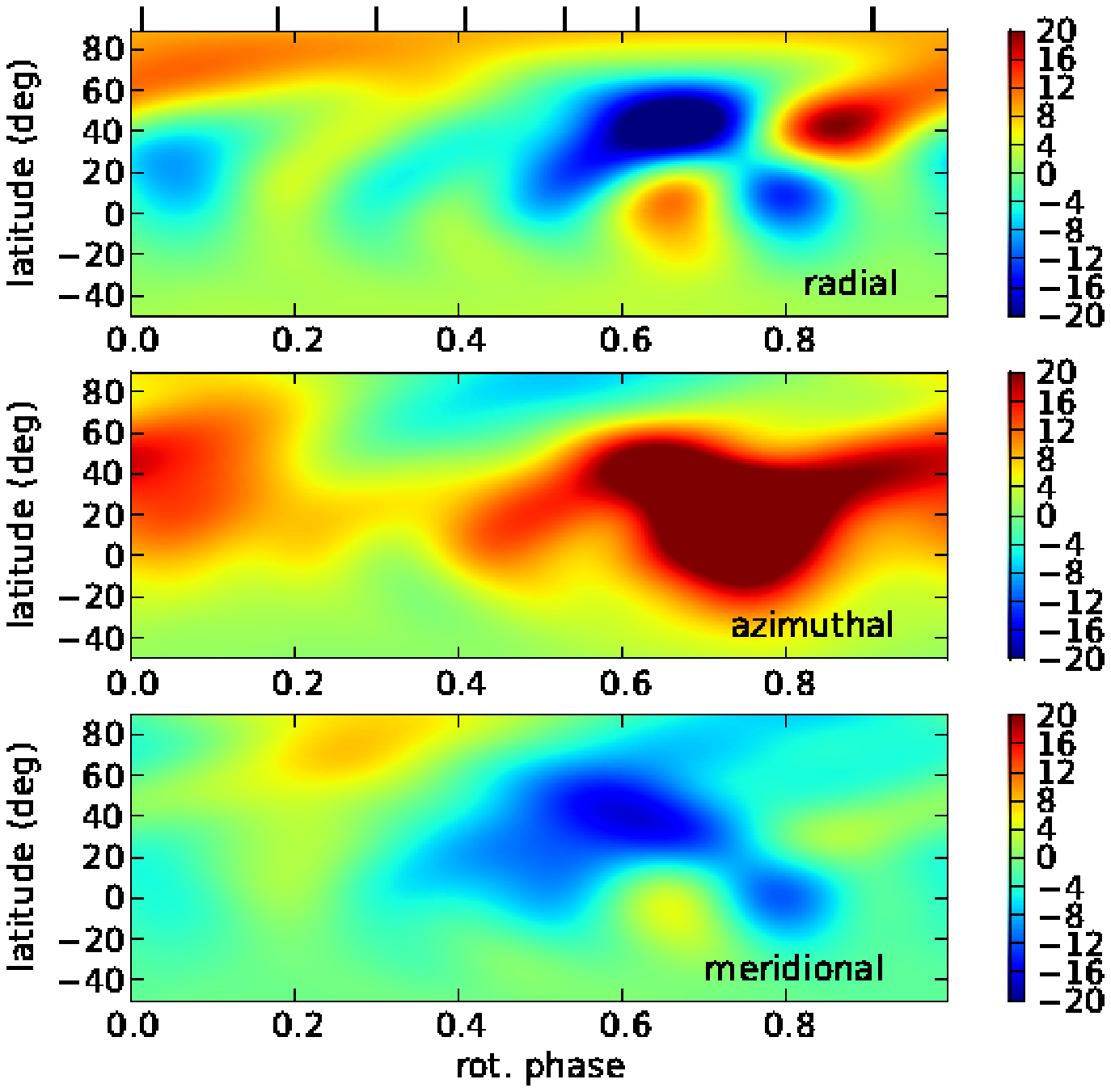}
}
\mbox{
\includegraphics[width=6cm]{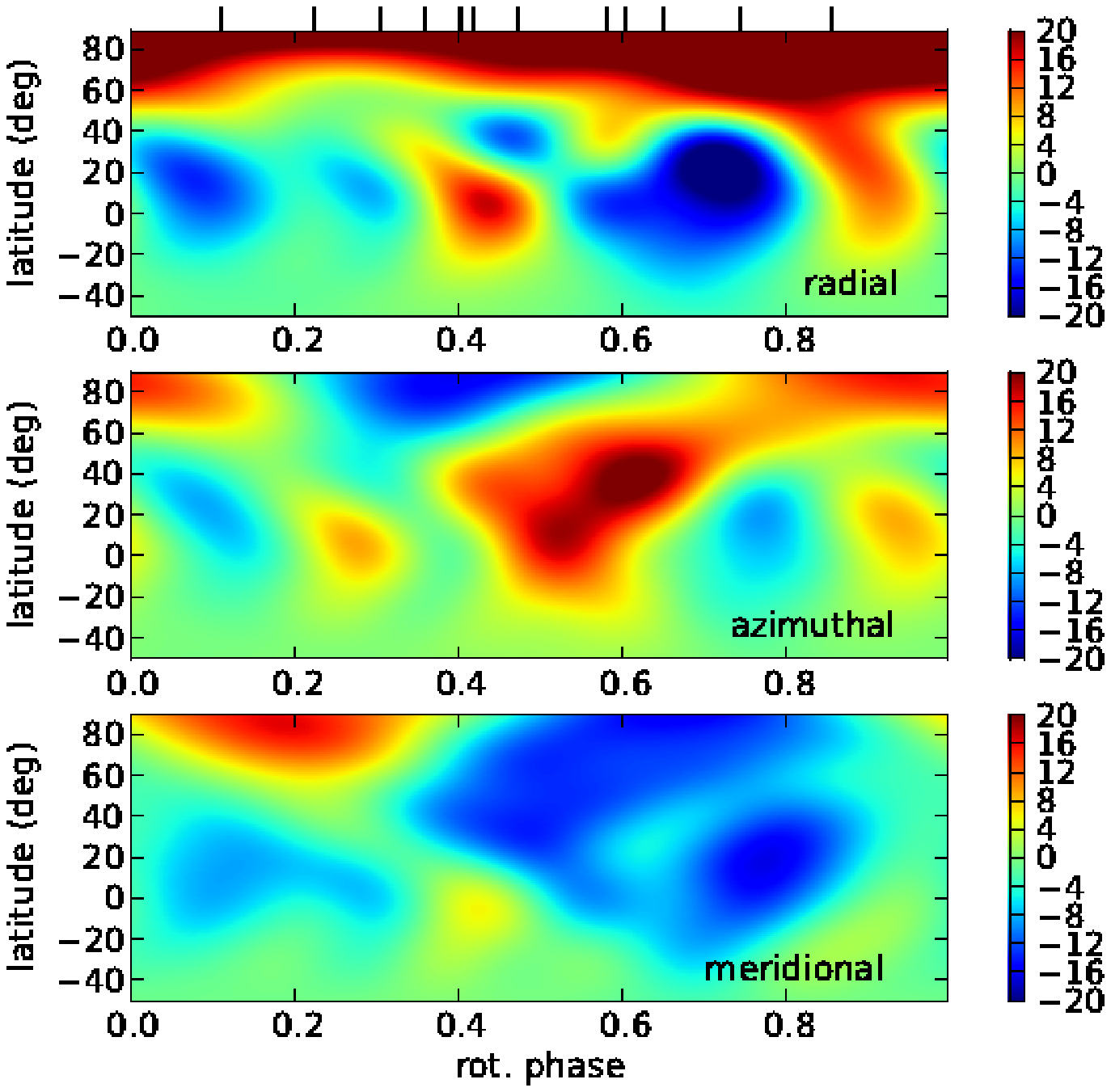}
\includegraphics[width=6cm]{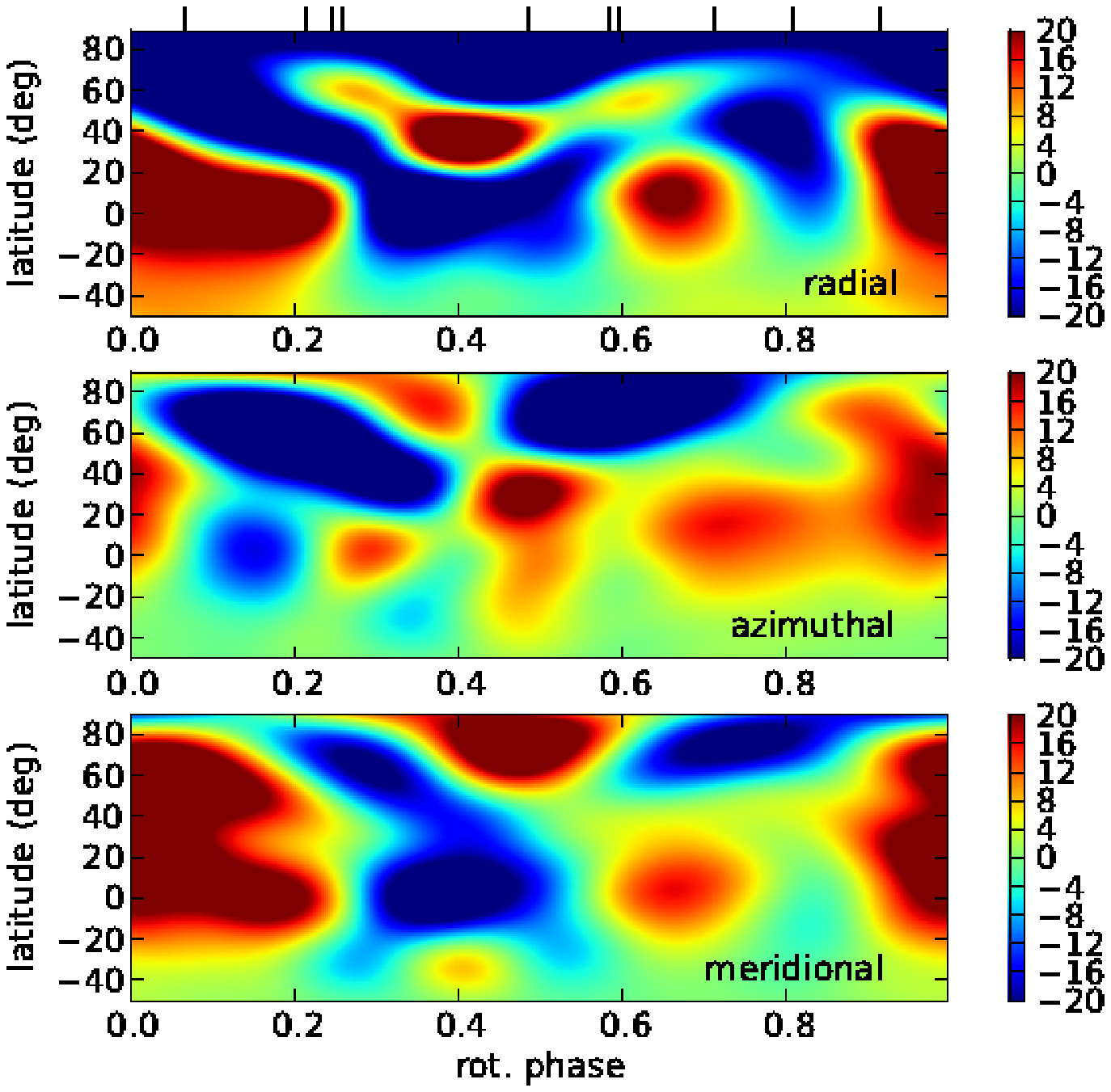}
}
\caption{Same as Fig. \ref{fig:hd78366} for HD 190771, for 2007.59, 2008.67, 2009.47 and 2010.50 data sets (from left to right and top to bottom).}
\label{fig:hd190771}
\end{figure*}

A more complex type of variability is illustrated by HD 190771. It has a mass similar to the Sun's, but has a rotation period of 8.8 days (Tab. \ref{tab:param}). In Fig. \ref{fig:hd190771}, we plot the magnetic maps derived for this star. A polarity reversal is visible on the strong azimuthal component between 2007.59 and 2008.67 (Petit et al. 2009). Between 2008.67 and 2009.47, the magnetic geometry changed in a different manner : the magnetic field which was mainly toroidal in 2008.67 became mostly poloidal in 2009.47. A second polarity reversal took place between 2009.47 and 2010.50, this time on the radial field component. In this case, the two successive polarity switches do not imply that the initial magnetic state is reached again, so that the observed variability is not taking the form of a cycle. 

In addition, we observe that the magnetic field intensity is correlated with the chromospheric emission. In the first three years, both the field strength and the chromospheric flux are roughly stable. We then observe a significant increase of these quantities in 2010.50 (Tab. \ref{tab:indice}).

\subsection{Fast and complex variability : $\xi$ Bootis A}

\begin{figure*}
\centering
\mbox{
\includegraphics[width=6cm]{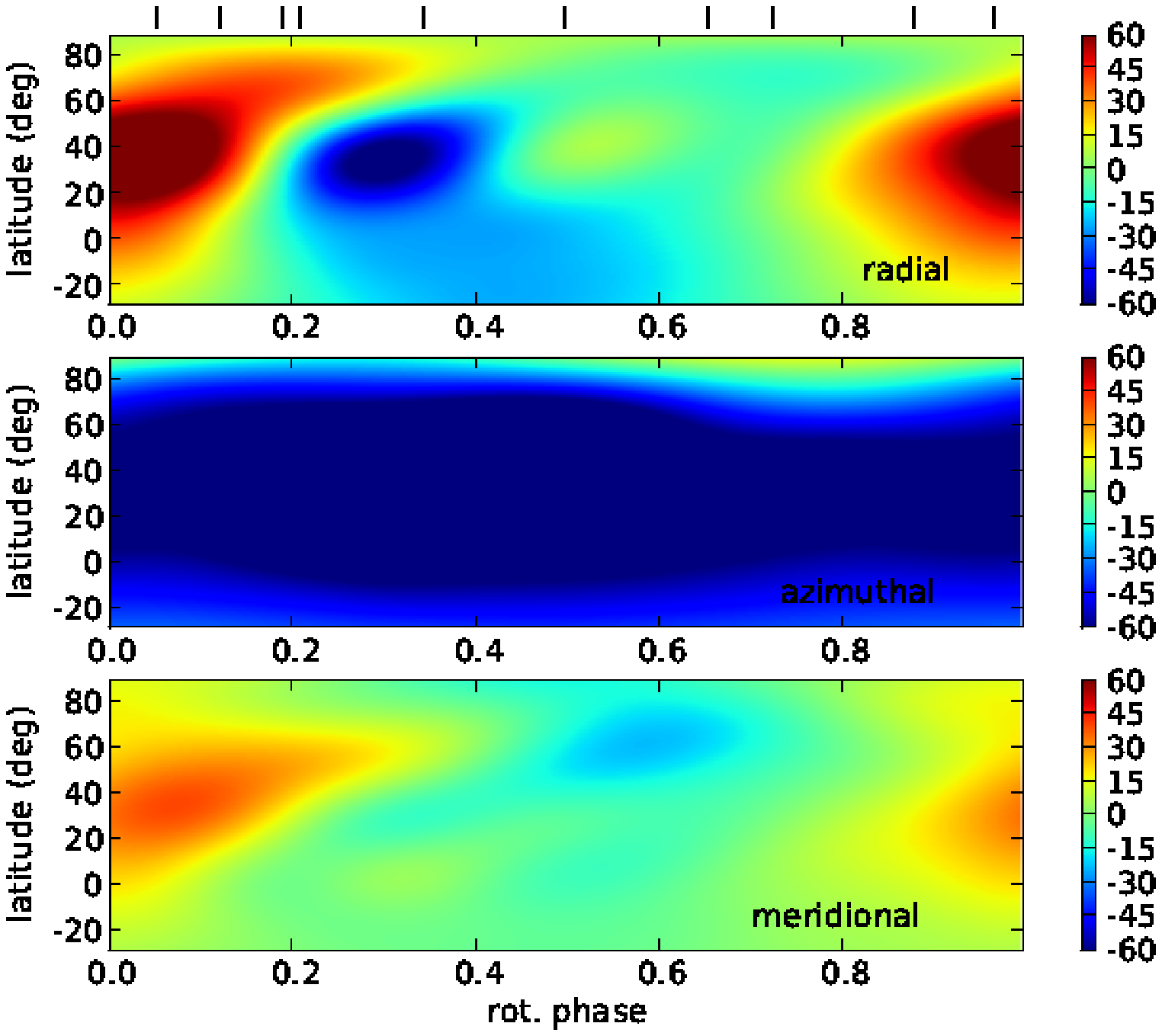}
\includegraphics[width=6cm]{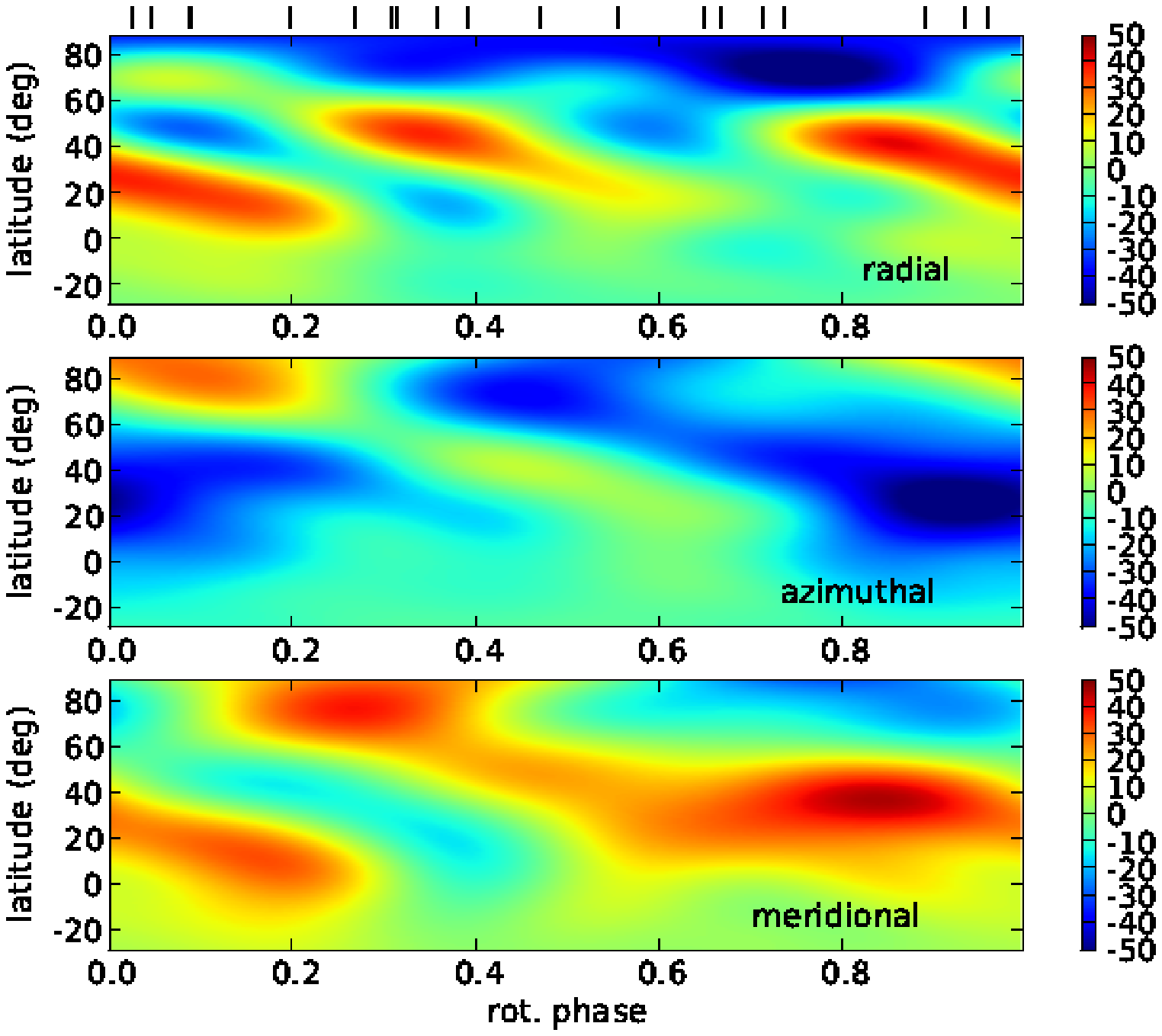}
}
\mbox{
\includegraphics[width=6cm]{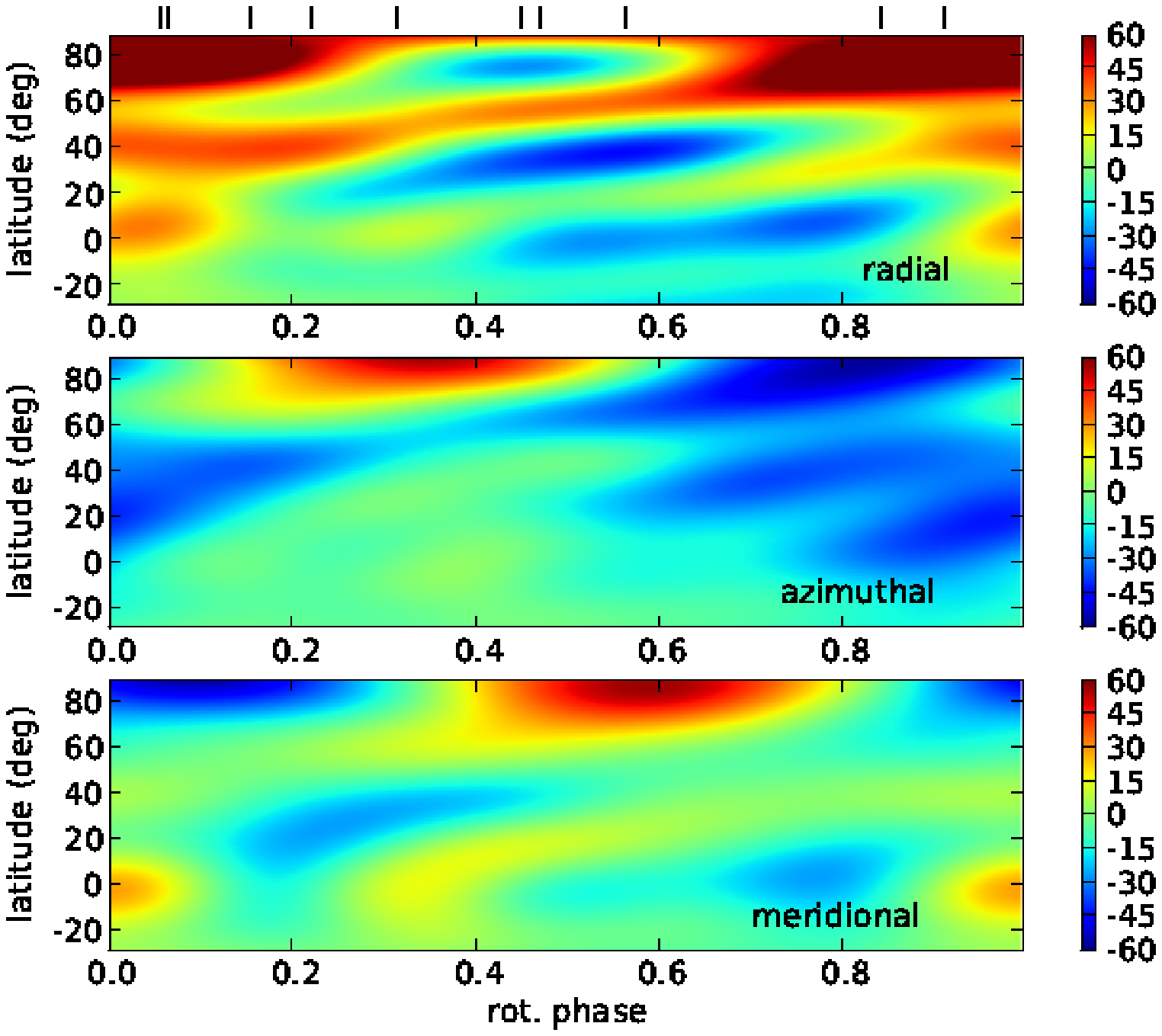}
\includegraphics[width=6cm]{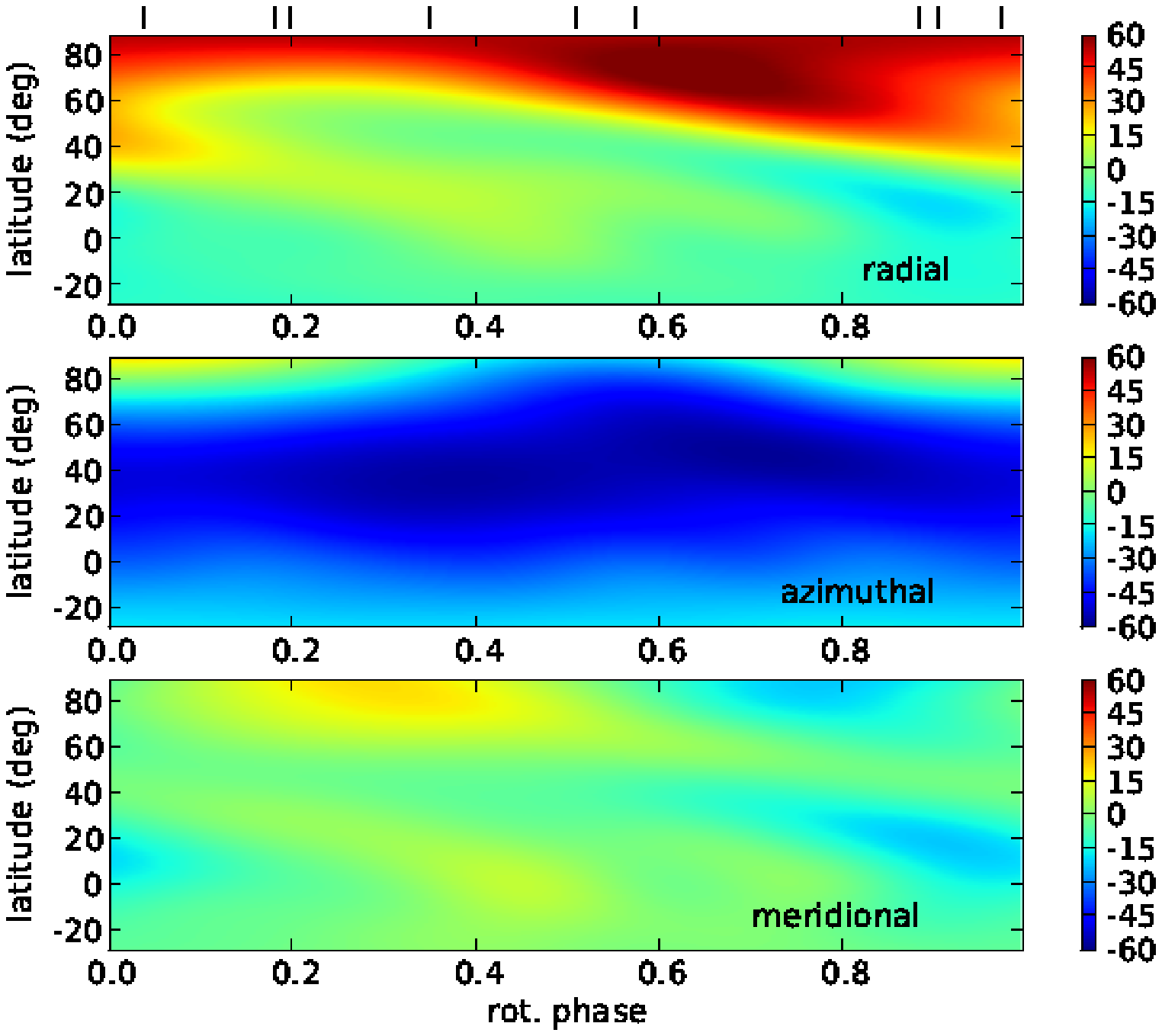}
}
\caption{Same as Fig. \ref{fig:hd78366} for $\xi$ Bootis A, for 2007.59, 2008.09, 2010.48 and 2010.59 data sets (from left to right and top to bottom).}
\label{fig:ksiboo}
\end{figure*}

Finally, another, more complex type of variability is observed with $\xi$ Boo A, the less massive and most rapidly rotating star of our three examples (Tab. \ref{tab:param}). It was observed at seven epochs, for which the magnetic field geometry and $N_{CaII H}$ were derived (Morgenthaler et al., submitted). Here we highlight two results of this long-term monitoring. 

The first one refers to the 2007.59 and 2008.09 data sets (top part of Fig. \ref{fig:ksiboo}). We observe that within a six months interval, the intensity of the magnetic field decreased by about 50\% and that the magnetic geometry, which was quite simple in 2007.59 with an aligned dipole and a prominent ring of azimuthal field, became more complex and less axisymmetric in 2008.09, with a less pronounced toroidal surface component. The decrease of the large-scale field strength is also observed in the chromospheric flux, with a sharp drop of emission between the two epochs (Tab. \ref{tab:indice}).

The second example is visible in the set of observations collected during the summer of 2010, which we decided to split in two subsets (2010.48 and 2010.59) to take into account the fast variations of the Zeeman signatures over this short timespan. In the correponding magnetic maps (bottom part of Fig. \ref{fig:ksiboo}), the most striking evolution is a sharp increase of the azimuthal magnetic field. These changes are taking place at a roughly constant level of chromospheric emission.

$\xi$ Boo A is therefore submitted to fast and complex surface changes that are different from those of the two previous stars, and reminiscent of the complex behaviour of other rapid rotators observed in the past (e.g. Donati et al. 2003).

\section{Discussion}

\begin{figure}
\centering
\includegraphics[scale=0.45]{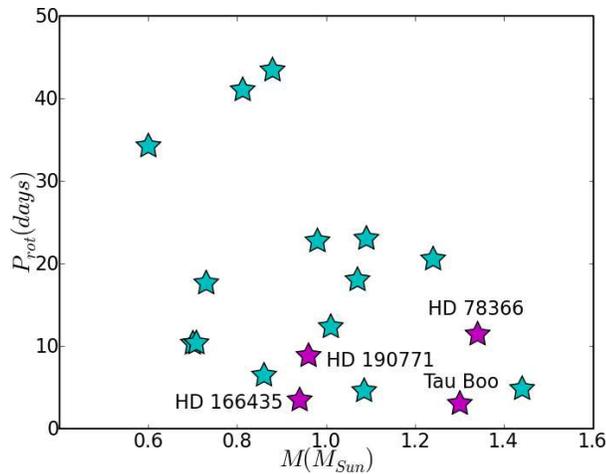}
\caption{Rotation period versus mass for the stellar sample. Pink symbols stand for stars with at least one polarity switch.}
\label{fig:m_prot}
\end{figure}

All stars of our sample show variability over the four years of our monitoring, but of different types. Stars which show at least one field reversal over this timespan have in common a fast rotation period (at least twice the solar one) and masses equal or slightly larger than that of the Sun (Fig. \ref{fig:m_prot}). In Fig. \ref{fig:m_prot}, we include $\tau$ Bootis, which is not part of our sample but which is reported to be affected by a short magnetic cycle of two years at most (Fares et al. 2009). We stress also that active stars with masses below our lower mass boundary (in particular, mid-M dwarfs with masses just below the fully convective limit) are reported to possess strong, simple and stable surface magnetic fields (Morin et al. 2008a,b).

$\tau$ Boo and HD 78366 were also observed at Mount Wilson as chromospherically active stars. For $\tau$ Boo, Baliunas et al. (1995) report a cycle of twelve years, versus two years from spectropolarimetry. Concerning HD 78366, periods of six and twelve years were identified using the Mount Wilson time-series, against about three years in our investigation. We therefore note that, at least for these two examples, the cycle lengths derived by chromospheric activity seem to be longer than those derived by spectropolarimetry. We suggest that this apparent discrepancy may be linked to the different temporal sampling inherent to the two approaches, so that the sampling adopted at Mount Wilson may not be sufficiently tight to unveil short activity cycles. 

Future observations of our stellar sample will allow us to investigate longer timescales of the stellar magnetic evolution. The sample includes several solar twins (Petit et al. 2008) which have not showed cycles yet, and which will help us to determine how small departures from the solar fundamental parameters may affect the characteristics of magnetic cycles. More generally, a regular monitoring of our targets over more than one decade will enable us to determine more precisely the relation between the length/oc-currence of magnetic cycles and the rotation/mass of Sun-like stars.

\acknowledgements
This research made use of the POLLUX data\-base (http://pollux.graal.univ-monpt2.fr) operated at LUPM (Universit\'e Montpellier II - CNRS, France, with support of the PNPS and INSU). We are grateful to the staffs of TBL for their efficient help during the many nights dedicated to this observing project.  

\newpage

\end{document}